\documentclass[letterpaper,english,aps, prl, twocolumn,longbibliography,showkeys]{revtex4-2}
\usepackage[T1]{fontenc}
\usepackage{verbatim}
\usepackage{textcomp}
\usepackage{amsmath}
\usepackage{amssymb}
\usepackage{graphicx}
\graphicspath{{fig/}}
\usepackage{bbold}
\usepackage{nicematrix}
\usepackage[colorlinks = true,citecolor=blue,linkcolor=blue,urlcolor=blue]{hyperref}

\newcommand{\dg}{^{\dagger }}
\newcommand{\bk} {{\bf k}}
\newcommand{\bR} {{\bf R}}
\newcommand{\br} {{\bf r}}
\makeatother

\usepackage{babel}
\begin{document}
\title{Triplet pairing mechanisms from Hund's-Kondo models:\\
applications to UTe$_{2}$ and CeRh$_{2}$As$_{2}$}
\author{Tamaghna Hazra$^1$}
\email{tamaghna.hazra@physics.rutgers.edu}
\author{Piers Coleman$^{1,2}$}
\affiliation{$^1$Center for Materials Theory, Rutgers University, Piscataway, New Jersey, 08854, USA\\
$^2$Department of Physics, Royal Holloway, University of London, Egham, Surrey TW20 0EX, UK}
\date{\today}
\begin{abstract}
Observing that several U and Ce based candidate triplet
superconductors share a common structural motif, with 
pairs of magnetic atoms separated by an
inversion center, we hypothesize
a triplet pairing mechanism based on an interplay of Hund's
and Kondo interactions that is unique to this structure. 
In the presence of Hund's interactions, 
valence fluctuations 
generate a triplet superexchange between electrons and 
local moments. The offset from the center of symmetry
allows spin-triplet pairs formed by the resulting Kondo effect 
to delocalize onto the Fermi surface, precipitating superconductivity.
We demonstrate this mechanism within a minimal two-channel Kondo lattice model
and present support for this pairing mechanism from existing experiments.
\end{abstract}
\maketitle

Hund's coupling is a direct manifestation of the Coulomb interaction
in multi-electron atoms.
One well-known consequence is the development
of large moments in $d$ and $f$-shell materials.
The effects of Hund's interactions on valence 
fluctuations and the Kondo effect in such high-spin systems 
are less well understood.

\begin{figure}[tb]
\includegraphics[width=0.8\columnwidth]{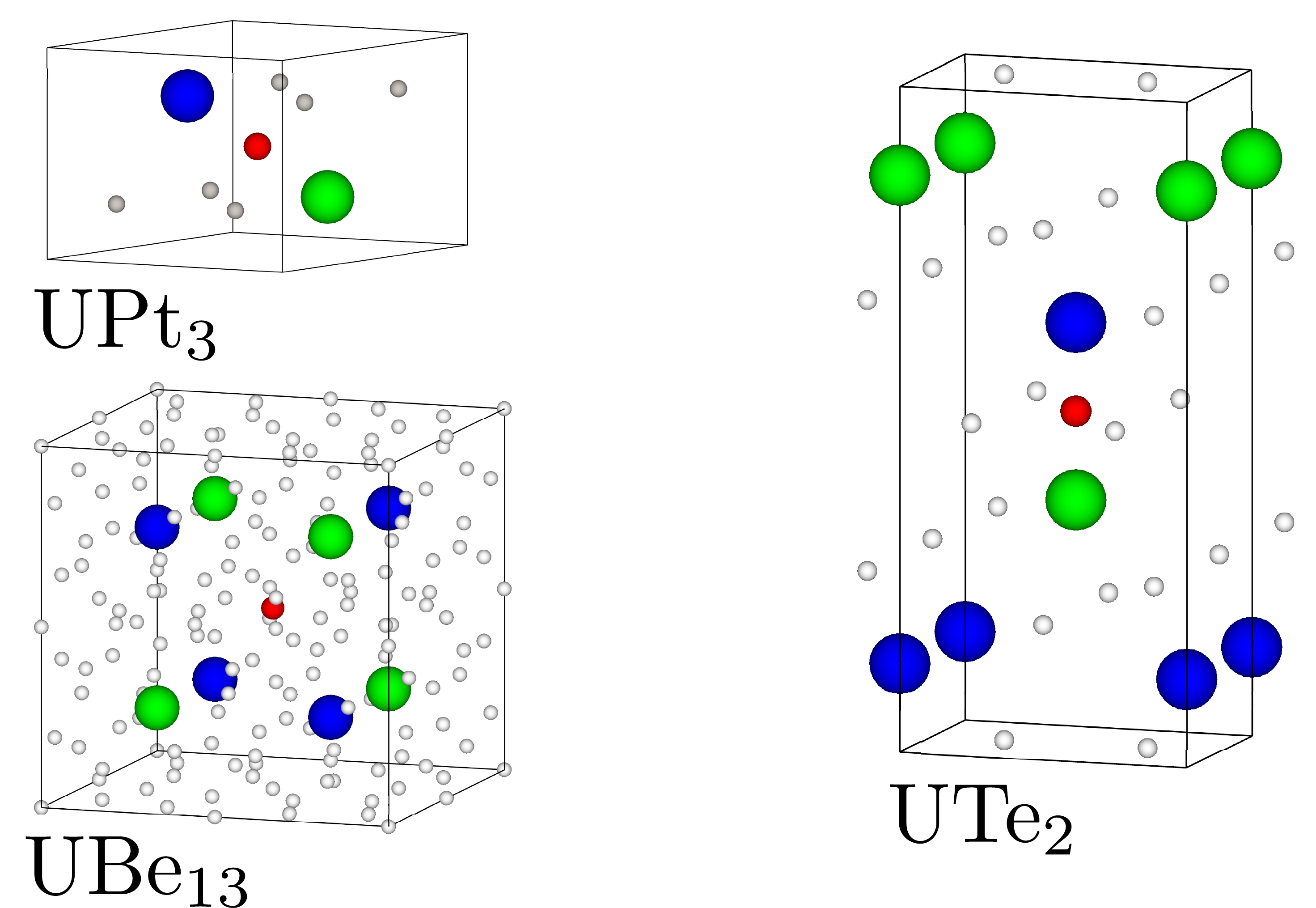}\caption{
Two-sublattice structure of 
UPt$_3$, UBe$_{13}$ and UTe$_2$, shown in the 
conventional
unit cell.
The inversion center is shown in red, and the two distinct sublattices
of U in blue and 
green.
\label{fig:UbasedStructs} }
\end{figure}

A possible role of Hund's coupling
in pairing mechanisms for triplet superconductivity has been discussed
by Anderson~\cite{anderson1985}, Norman~\cite{norman1994}, Hotta and Ueda~\cite{hotta2003,hotta2004} and more
recently in the context of triplet resonating valence bonds (tRVBs) in various
settings~\citep{coleman2020,konig2021}. 
Here we discuss the implication of these ideas for triplet-paired
heavy fermion superconductors in a two-channel Kondo lattice model~\cite{schiller1998,cox2020,coleman1999,flint2008,flint2011}, with two key observations.
First, we highlight a common structural motif unique to candidate triplet-paired heavy fermion materials (Table	
\ref{tab:Heavy-fermion-triplet} and Figure \ref{fig:UbasedStructs}) - 	
an even number of magnetic ions	
separated by a center of symmetry. This sublattice structure enables the triplet	
paired correlations induced by Hund's coupling to develop coherence	
on the Fermi surface, inducing a Cooper instability. 	
Second, we show how Hund's coupling modifies 	
the structure of the Kondo interaction, inducing	
triplet correlations between  scattered electrons and local	
moments. 
\setlength{\arraycolsep}{0.8pt}
\renewcommand{\arraystretch}{1.0}
\definecolor{gray}{gray}{0.95}
\newcommand{\mc}[2]{\multicolumn{#1}{c}{#2}}
\begin{table}[b]
%
\resizebox{\columnwidth}{!}{
\begin{tabular}{ |l|c|c|c|c|c|c|c|c|}
\hline  
 & UTe$_{2}$ & UGe$_{2}$ & UCoGe 
 & URhGe 
 & UBe$_{13}$
 &UPt$_{3}$&\small{CeRh$_{2}$As$_{2}$}&CeSb$_2$
\\
\hline
Ref&
\cite{aoki2021}
&\cite{aoki2019a} 
&\cite{aoki2019a} 
&\cite{aoki2019a} 
&\cite{chen1985}
&
\cite{dale2009}
&\cite{khim2021} 
&\cite{squire2022} 
\\
\hline
n$_{M}$&2&2&4&4&2&2&2&2
\\
\hline
 T$_{c}$& $2$K & $0.8$K & $0.8$K 
 & $0.25$K 
 & $0.95$K & $0.5$K & $0.26$K &$0.22$K\\
\hline 
H$_{P}$ & $3.7$T & $1.5$T & $1.5$T 
&$0.5$T 
& $1.8$T & $0.9$T & $0.5$T & $0.4$T\\
\hline 
H$_{c2}$ & $60$T & $3$T & $18$T 
& $13$T 
& $14$T & $2.8$T & $14$T & $3$T\\
\hline 
T$_{M}$ & - & $52$K & $2.7$K 
& $9.5$K 
& - & - & - & -\\
\hline 
\end{tabular}}
\caption{\textbf{Candidate heavy fermion triplet superconductors}
have either n$_{M}=$2 or 4 magnetic U/Ce atoms
in the conventional unit cell separated by an inversion center.
Maximum superconducting
T$_{c}$, Pauli limiting critical field H$_{P}$, highest measured
upper critical field H$_{c2}$, Curie temperature T$_{M}$ for ferromagnetic
superconductors.
\label{tab:Heavy-fermion-triplet}}
\end{table}

\begin{figure}
\includegraphics[width=0.9\columnwidth]{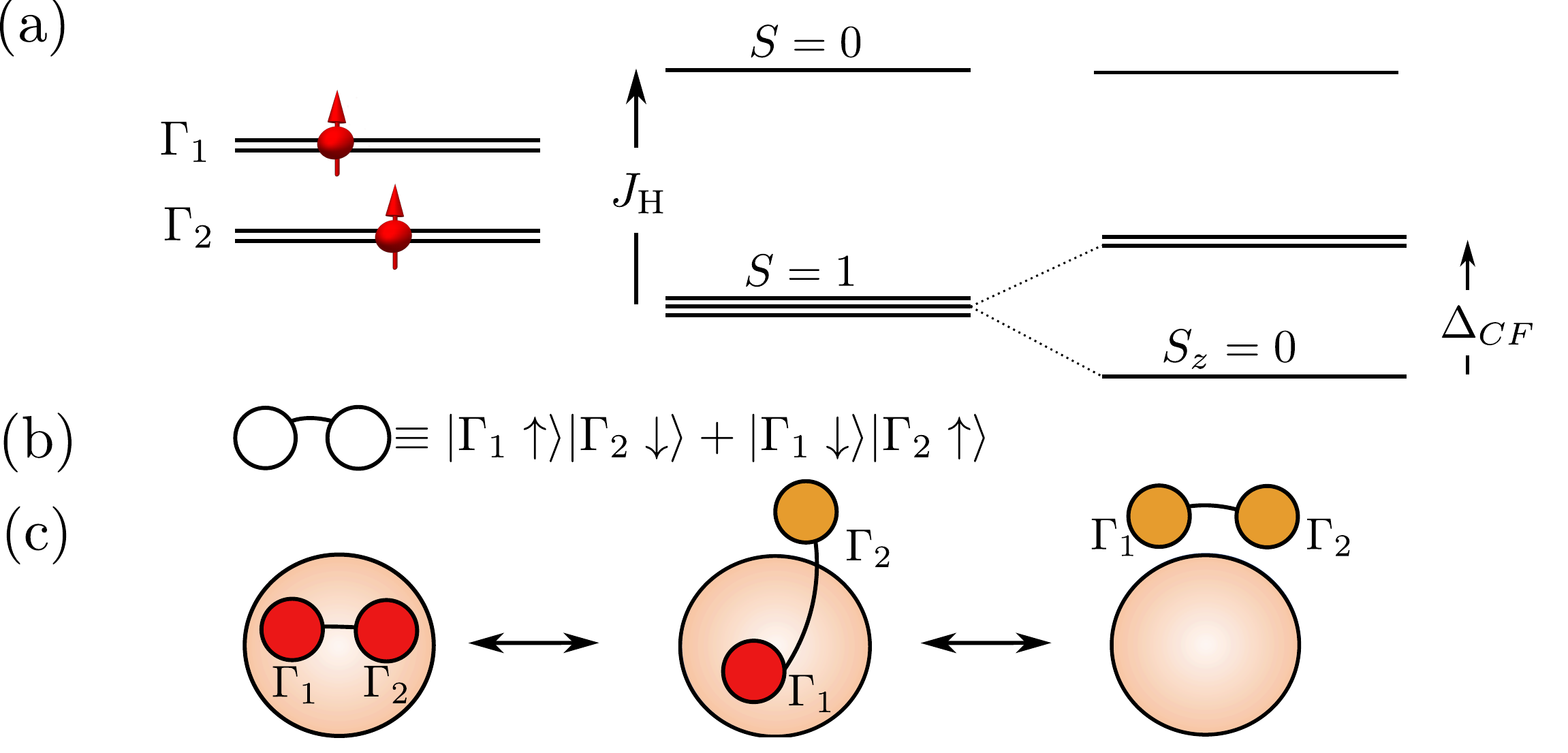}
\caption{\textbf{Delocalization of Hund's coupled triplet pairs: }
(a) Hund's coupling ($J_{H}$) and crystal field ($\Delta_{CF}$) 
entangle two spins in an $f^{2}$ local moment 
into an $S_{z}=0$ triplet state. (b) 
Triplet valence bond representation of $S_{z}=0$ triplet state. (c) 
Hybridization causes a transfer of triplet pairs (red) into the 
conduction (orange) sea. 
\label{fig:Triplet-pairing-driven}}
\end{figure}

Our observations are built on the tRVB concept~\citep{coleman2020,konig2021}, whereby
valence fluctuations into entangled, high-spin configurations
(Figure \ref{fig:Triplet-pairing-driven}) can
delocalize into a coherent triplet RVB superconducting state. 
Symmetry plays a central role in this process, 
for triplet-paired configurations in the excited state of a magnetic atom
are even
under inversion about the atom, while triplet pairing on a 
Fermi surface is necessarily odd parity. Anderson~\citep{anderson1985}
recognized that if local moments are situated at distinct sublattices
away from the inversion centers, the onsite triplet pairs could 
acquire an inversion-odd sublattice form factor, allowing them to coherently
couple to triplet Cooper pairs on the Fermi
surface~\citep{norman1994,hotta2003,hotta2004}.  
Remarkably, the structural motif identified by Anderson forty years ago characterizes every candidate triplet-paired heavy fermion superconductor we know today, with only two exceptions~\cite{nguyen2021},\footnote{UAu$_2$ also seems to violate the Pauli limit, Andrew Huxley et al, PNAS in press (2022)}. 
(see Table \ref{tab:Heavy-fermion-triplet}
 and Figure \ref{fig:UbasedStructs}).
Recently,
CeRh$_{2}$As$_{2}$ and CeSb$_2$ have also emerged as new members~\citep{khim2021,squire2022}
of this class of compounds
\footnote{As pointed out in the
Supplementary of \citep{khim2021}, an
odd-parity spin-triplet order parameter with the d-vector along $z$ leads to the same pseudo-spin triplet pairing
order parameter on the Fermi surface as an odd-parity spin-singlet
pairing order parameter.}.
The near-universal correlation suggests the possibility of another family of triplet superconductors in PrTr$_2$Al$_{20}$ (Tr=Ti,V)~\citep{sakai2011,sakai2012}, which also has this structural motif~\citep{okuyama2019}.

{\color{black}
To illustrate the interplay between Hund's coupling and valence
fluctuations, consider a magnetic  ion in which the
f-electrons exist in three  valence
states, taken for simplicity to be 
f$^{0}$, f$^{1}$ and f$^{2}$. Suppose that in the f$^{2}$ configuration,
Hund's coupling and crystal-field splitting, enabled by spin-orbit coupling, stabilize a state in which two f-orbitals of 
symmetry $\Gamma_{1}$ and $\Gamma_{2}$ are entangled into an $S=1$,
$S_{z}=0$ state (Fig. 2a), forming a triplet valence bond between
the two orbitals (Fig. 2b). 
When the ion hybridizes with the conduction sea, valence
fluctuations will now allow the 
entangled triplet pair to escape into the conduction sea (Fig. 2c).

To understand how Hund's interactions affect super-exchange,
suppose the 
f$^{1}$ $\Gamma_{1}$ Kramer's configuration is the most stable, forming a local moment. 
\begin{figure}[b]
\includegraphics[width=1\columnwidth]{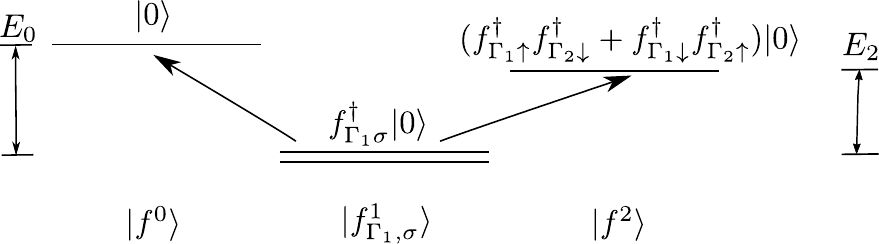}\caption{\textbf{
Low energy Hilbert space of the local moment, showing associated
Hubbard operators: } 
$\Gamma_{1}$ is the irreducible representation
of the spatial wavefunction of the ground-state  $f^{1}$ doublet. The f$^{2}$ excited stated is a
entangled spin-triplet state with $S_{z}=0$, formed from electrons in
the $\Gamma_{1}$ and $\Gamma_{2}$ orbitals. 
}
\label{fig3}
\end{figure}
Integrating out the virtual valence fluctuations~(Fig. \ref{fig3}), ${\rm f}^{1}\rightleftharpoons {\rm f}^{0}+e^{-}$ and $e^{-}+{\rm f}^{1}\rightleftharpoons {\rm f}^{2}$ via a Schrieffer-Wolff transformation~\cite{Schrieffer66,Coqblin69} generates a
second-order perturbation in the 
energy of conduction electrons scattering in the $\Gamma_{1,2}$ channel
\begin{equation}\label{}
H_{K} = - \frac{|V_{\Gamma_{1}}|^{2}}{E_{0}}P^{\Gamma_{1}}_{S} - \frac{| V_{\Gamma_{2}}|^{2}}{E_{2}}P^{\Gamma_{2}}_{T}
\end{equation}
where  $V_{\Gamma_{1,2}}$ are the hybridization matrix elements in the two
channels, while $E_{0}$ and $E_{2}$ are the corresponding excitation
energies. 
The operators
$P^{\Gamma_{1}}_{S}= \frac{1}{2} - \vec{\sigma
}^{\Gamma_{1}}\cdot \vec{S} 
$  and 
$P^{\Gamma_{2}}_{T}= \sigma_{z}P^{\Gamma_{2}}_{S}\sigma_{z}$ project
the incoming quasiparticles 
into the singlet and triplet states of the excited f$^{0}$ and
f$^{2}$ states, respectively.  
Here $\vec{\sigma}^\Gamma$ is the spin of the conduction electron in channel $\Gamma$.
By noting that a $S_{z}=0$ triplet 
$\mid\Downarrow \uparrow+\Uparrow\downarrow\rangle=\sigma_{z} \mid\Downarrow \uparrow- \Uparrow\downarrow \rangle$
is obtained from a 
singlet 
by rotating the conduction electron spin
through 180$^{\circ}$ about the $z-$axis, we have written  $P^{\Gamma_{2}}_{T}=
\sigma_{z}P^{\Gamma_{2}}_{S}\sigma_{z}$ as a unitary transform of the
singlet operator $P^{\Gamma_{2}}_{S}$.

Omitting  potential scattering
terms, it follows that Hund's interactions cause the 
Kondo interaction to 
develop a triplet-superexchange with XXZ anisotropy,
\begin{eqnarray}\label{}
H_{K} &=& J_{1}{\bf S}\cdot \boldsymbol{\sigma }^{\Gamma_{1}}+
J_{2} {\bf S}\cdot (\sigma_z \boldsymbol{\sigma}^{\Gamma_{2}}
\sigma_z)
\end{eqnarray}
where $J_{1} = |V_{\Gamma_{1}}|^{2}/ (2E_{0})$ and 
$J_{2} = |V_{\Gamma_{2}}|^{2}/ (2E_{2})$~\footnotemark[22].  The 
Hund's-Kondo term can alteratively be written 
$H_{K2}=J_{2}[ S_{z}\sigma^{\Gamma_{2}}_{z}-
S_{x}\sigma^{\Gamma_{2}}_{x}- S_{y}\sigma^{\Gamma_{2}}_{y}]$.
Acting in isolation (\textit{i.e.} if $J_{1}=0$)
the Hund's-Kondo coupling $J_{2}$ flows to strong-coupling
like its antiferromagnetic counterpart~\cite{anderson1971},
but forms a ``screened'' triplet state
($\mid\Uparrow\downarrow +\Downarrow \uparrow\rangle $). We are
interested in the interplay of the two terms in the lattice. 

Although we have chosen an $S_{z}=0$ orientation of the
Hund's triplet to illustrate this physics, 
in practice, the crystal fields 
will determine the orientation of the d-vector
of the triplet f$^2$ excited state. 
Moreover, spin-orbit coupling will generically 
introduce additional rotations of the electron spin-quantization axis
into the hybridization matrix elements. These two effects mean that 
pre-formed triplet pairs of any 
odd-parity irreducible representation allowed by the crystal structure
can delocalize via the Kondo
hybridization. 

We now incorporate the above effects into a two channel 
Kondo lattice model $H=H_{c}+H_{K1}+ H_{K2}$, where 
\begin{eqnarray}\label{eq:Hamsplit}
H_{c}&=& - \sum_{\mathbf{k}}c\dg_{\bk
}
\bigl[(t_{0}\gamma_\bk+\mu) +
t_{1}\gamma_\bk\alpha^{x}\bigr]c_{\bk},\cr
H_{K1}&=&  J_{1}\sum_{j\alpha}
\psi\dg _{1\alpha} (j)\boldsymbol{\sigma}\psi_{1\alpha}(j)
\cdot\mathbf{S}_{j\alpha},\cr
H_{K2}
&=&  J_{2}\sum_{j\alpha}
\psi\dg _{2\alpha} (j)\sigma_{z}\boldsymbol{\sigma}\sigma_{z}\psi_{2\alpha}(j)
\cdot\mathbf{S}_{j\alpha}.
\end{eqnarray}
Here $H_{c}$ describes electron hopping on a 
two-sublattice body-centered 
cubic lattice,
reminiscent of 
UTe$_2$~\citep{ishizuka2021}, where $c\dg_{\bk }\equiv c\dg_{\bk \alpha \sigma }$ creates an electron of wavevector
$\bk $ on sublattice $\alpha =\pm 1$ with spin component $\sigma^{z}=\sigma
$,
$t_0$ and $t_{1}$ are the intra- and inter-sublattice hopping integrals, respectively,
$({\alpha}^{x},{\alpha}^{y},{\alpha}^{z}) $ are the sublattice
Pauli matrices and
$\gamma_\mathbf{k}=8\cos\frac{k_{x}}{2}\cos\frac{k_{y}}{2}\cos\frac{k_{z}}{2}$ 
is the nearest neighbor form-factor that is invariant under the D$_{2h}$ point group.
$H_{K1}$ and $H_{K2}$ are 
the Kondo interaction in channels $\Gamma= (\Gamma_{1},\Gamma_{2})$, 
where 
\begin{equation}\label{}
\psi_{\Gamma \alpha}^\dagger (j) = \frac{1}{\sqrt{N_{s}}}\sum_{\bk }
c\dg _{\mathbf{k}\alpha} \Phi_{\Gamma\bk }e^{-i \bk \cdot \bR_j}
\end{equation}
create electrons in Wannier states of symmetry $\Gamma$ coupled to
spins $\mathbf{S}_{j\alpha }$ at site $j,\alpha $ ($N_s$ is the number
of unit cells).  We choose $\Phi_{1\bk}=1$ and 
\begin{equation}\label{}
\Phi_{2\bk }=
i\sigma_{x}(\sqrt{1-\zeta^{2}} +i\zeta \alpha^{z}p^{z}_{\bk}), 
\end{equation}
for
the two Kondo channels, consistent with time-reversal and inversion
symmetry. Here $p^{z}_{\bk } = \cos (\frac{k_{x}}{2}) \cos
(\frac{k_{y}}{2}) \sin (\frac{k_{z}}{2})$ is an odd-parity crystal
harmonic transforming under the $B_{1u}$ representation.  The
coefficient $\zeta$ is finite when the two magnetic atoms are
displaced from their common center of symmetry, activating
an antisymmetric spin-orbit~\citep{bauer2012} mediated
coupling between tRVBs (Fig. \ref{fig:Triplet-pairing-driven}b) and
triplet Cooper pairs. 
The factor $i\sigma_x$ in the hybridization 
$\Phi_{2\bk }$ captures the spin-orbit
coupling between 
f-states and a
conduction state with different orbital content. The term
proportional to $\zeta$
describes a coupling between spin and momentum
that is odd in the sublattice index~\footnotemark[22], similar to the
staggered Rashba coupling~\citep{sigrist2014,ishizuka2021} discussed
in the context of layered materials like CeRh$_2$As$_2$.

The key physics of this model involves a co-operative
action of the Kondo effect in the two channels~\cite{coleman1999,flint2008,flint2011}. 
At high temperatures in the lattice, 
the Kondo coupling in both channels renormalizes to strong-coupling
according the scaling equation $\partial
(J_{\Gamma}\rho)/\partial\ln \Lambda = - (J_{\Gamma}\rho)^{2}$~\cite{anderson1970a},
where $\rho $ is the conduction electron density of states,
and $\Lambda$ the energy cut-off. 
Suppose channel one is the strongest channel with the largest Kondo
temperature $T_{K1}\sim D e^{-1/ J_{1}\rho
}$, where $D$ is the bandwidth. Then the logarithmic renormalization in
channel two is interrupted at a scale $\Lambda=T_{K1}$, with a renormalized 
Kondo coupling constant given by
\begin{equation}\label{coupling}
\frac{1}{J_{2}^{*} } =  \frac{1}{J_{2}^{0}}- \rho \ {\ln \left (\frac{D}{T_{K1}} \right)} 
\end{equation}
At temperatures $T\lesssim T_{K1}$ the 
local moments fractionalize into deconfined heavy fermions
$\mathbf{S}_{j} \rightarrow f\dg_{j}
\left(\frac{\mathbf{\sigma }}{2} \right)
f_{j}$:
in an impurity model, a Kondo resonance forms in the strongest channel
and nothing  further happens. 

However, in the 
Kondo lattice, the heavy fermions hybridize with the conduction
electrons to  form a large Fermi surface~\cite{oshikawa2000,hazra21}. 
Moreover, in the presence of a finite $\zeta$ 
the residual interaction $J_{2}$ created by valence fluctuations into 
Hund's-coupled spin-triplet states, couples triplet Cooper pairs
on the Fermi surface, which reactivates its scaling, causing it 
to resume its upward logarithmic renormalization, 
ultimately diverging at $T_{c}$ to form a Hund's driven triplet superconductor. 

To examine this process, we note that action of
$H_{K2}$ on the deconfined fermions
is given by 
\begin{equation}\label{pairit}
H_{K2}= -J_{2}^{*}\sum_{j }
 (\psi_{2 j}\dg 
\sigma_{z} (-i\sigma_{y})f\dg_{j })
 (f_{j } (i\sigma_{y})\sigma_{z} 
\psi_{2j }) 
\end{equation}
where the sublattice indices have been suppressed.
Here we have used the particle-hole symmetry of
the fractionalized spin operator to replace $f_{j }\rightarrow
-i\sigma_{y}f\dg_{j }$ in the usual hybridized form of the Kondo
interaction. Now the hybridized quasiparticle operators formed in
channel one have the form $a_{\bk \sigma } = u_{\bk }f_{\bk  \sigma } + v_{\bk  }\Phi_{1\bk }c_{\bk  \sigma }$,
where $u_{\bk }\sim 1$, while $v_{\bk }^{2}\sim \frac{m}{m^{*}}\ll 1$ is set by the inverse mass
renormalization of the heavy fermions~\footnote{For simplicity, we have ignored the inter-sublattice hopping for now, and suppressed the sublattice indices in $f_{\bk  \sigma}$. In general, $f_{\bk\sigma\alpha},c_{\bk\sigma\alpha}$ must be replaced by the corresponding band eigenstates, and $u,v$ are matrices as shown explicitly in the Supplementary Material[22] Section II}. 
If we now decompose $H_{K2}$ as a pair scattering potential acting on the
quasiparticle pairs at the Fermi level, we obtain
\begin{equation}\label{pairscat}
H^{*} =- J_{2}^{*} |u_{\bk }v_{\bk }|_{\rm FS}^{2}\sum_{\bk, \bk '}\Lambda_{\bk }\dg \Lambda_{\bk' }
\end{equation}
where $|u_{\bk }v_{\bk }|_{\rm FS}^{2}$ denotes the Fermi
surface average of the coherence factors and $\Lambda_{\bk } = 
a_{-\bk } (i\sigma_{y})
(\mathbf{d}^{A}_{ \bk }\cdot\mathbf{\sigma })
a_{\bk}$ is the projection of the triplet pair operators
in \eqref{pairit} onto the Fermi surface, 
where
$\mathbf{d}^{A}_{\bk }=(\mathbf{d}_{\bk
}-\mathbf{d}_{-\bk })/2$ is the odd-parity 
component of the form factor
$\mathbf{d}_{\bk }\cdot\mathbf{\sigma }=
\alpha_z \sigma_{z}\Phi_{2\bk }\Phi_{1\bk }$, projected into the band eigenstates \footnotemark[22].
\footnotetext[22]{See Supplementary  Materials for details of the derivation of the Kondo Hamiltonian and projection to the heavy fermion bands}
This antisymmetric term is absent 
if the magnetic ions lie at a center of symmetry, both $\Phi_{1\bk }
$ and $\Phi_{2\bk }$ have the same parity under $k\to -k$, so
$\mathbf{d}^{A}_{\bk }=0$ and the triplet pseudopotential vanishes. 
However, in the presence of an offset center of symmetry, 
($\zeta>0$) it becomes finite.
In our model calculation, 
$\mathbf{d}^{A}_{\bk }\sim i \zeta  p^{z}_{\bk }\hat{\mathbf{y}}$
distinct from the d-vector of the localized f-electrons in the moment (Fig.~\ref{fig:Triplet-pairing-driven}), due to the spin-orbit coupling~\footnotemark[22].

From these arguments, we see that the triplet coupling constant
induced by the Hund's Kondo effect is now $g_{t}\sim 
J_{2}^{*} \rho^{*}(uv)^{2}_{FS}\approx J_{2}^{*} \rho^{*}v^{2}$
because $u_{\bk }\sim 1$.  Now at first sight, the small size of 
$v^{2}\sim m/m^{*}\ll 1$ is cause for concern, but it is compensated by the large density of states of the heavy Fermi liquid $\rho^*\sim 1/T_{K1}$.
In fact the product 
$v^{2}\rho^{*}= dN/d\mu$ is recognized as the 
charge susceptibility of the heavy
Fermi liquid, and since the 
f-component of the heavy Fermi liquid is incompressible, this quantity
is equal to the unrenormalized density of states of the conduction
fluid, $dN/d\mu\sim \rho $, so that $g_{t}\sim 
\zeta J_{2}^{*}\rho $ is essentially unaffected by the large mass
renormalization of the heavy electrons. 
A Cooper instability will develop when $\frac{1}{J_{2}^{*}}- \rho  
d^{2}\ln  \left(\frac{T_{K1}}{T_{c}} \right)=0$,
where we have denoted the average d-vector magnitude by $d^{2}= \langle |{\mathbf{d}}^{A}_{\bk
}|^{2}\rangle_{FS}$. Combining with \eqref{coupling}, a superconducting instability will
take place when 
\begin{equation}\label{Tceqn}
\frac{1}{J_{2}^{0}}- \rho \ {\ln \left (\frac{D}{T_{K1}} \right)}
-  \rho  d^{2}
\ln  \left(\frac{T_{K1}}{T_{c}} \right)=0
\end{equation}
Remarkably, this expression contains two logs - a Kondo log that
renormalizes $J_{2}$ from the band-width down to $T_{K1}$, followed by a further
Cooper renormalization of the coupling constant below $T_{K1}$.
Thus the second channel Kondo effect plays a vital 
co-operative role in enhancing the pairing process.
Solving \eqref{Tceqn}, we find that 
\begin{equation}\label{}
T_{c} = (T_{K2})^{\frac{1}{d^{2}}} (T_{K1})^{1-\frac{1}{d^{2}}}
\end{equation}
is a weighted
geometric mean of the Kondo temperatures $T_{K\Gamma}= D
e^{-1/J_\Gamma \rho}$ in the two channels, emphasizing the
co-operative nature of the Hund's-Kondo effect. 

To quantify this effect in greater detail, we employ an 
SU(2) gauge theory\citep{coleman1989,wen1996} description of the Kondo effect.
The local moment
on each sublattice may be represented in terms of Abrikosov
pseudo-fermions,
$\mathbf{S}=\frac{1}{2}(f^{\dagger}\boldsymbol{\sigma}f)$.
In the Nambu basis $F=\big(\begin{array}{cc}
f, & (i\sigma_{y})^{\dagger}f^{\dagger}\end{array}\big)^{T}$, this corresponds to $\mathbf{S}=\frac{1}{4}F^{\dagger}(\boldsymbol{\sigma}\otimes\tau_{0})F$,
a representation that is invariant under SU(2) particle-hole transformations
of the spinons $F\rightarrow e^{i\mathbf{n\cdot}\boldsymbol{\tau}}F$, where $\tau_i$ are Pauli matrices in Nambu space. 
In this representation, the Kondo couplings become four-fermion interactions
which are then decoupled using a hybridization mean-field $V_{\Gamma\alpha}$
and a `pairing' field $\Delta_{\Gamma\alpha}$ in each channel and sublattice to get
$H=H_{c}+H_{K}$, where
\begin{eqnarray}\label{}
H_{K} &=& \sum_{\Gamma \mathbf{k}\alpha}\left(
V_{\Gamma\alpha} 
\tilde{\psi}_{\Gamma \mathbf{k}\alpha}^{\dagger}
f_{\mathbf{k}\alpha}
+\Delta_{\Gamma\alpha}
\tilde{\psi}_{\Gamma \mathbf{k}\alpha}^{\dagger}
(i\sigma_{y})^{\dagger}
f_{-\mathbf{k}\alpha}^{\dagger}+\mathrm{H.c.}\right)\nonumber \\
&+& N_{s}\sum_{\Gamma\alpha}\frac{\left|V_{\Gamma\alpha}\right|^{2}+\left|\Delta_{\Gamma\alpha}\right|^{2}}{J_{\Gamma}}.
\end{eqnarray}
where $\tilde{\psi}_{\Gamma_1}=\psi_{\Gamma_1}, \tilde{\psi}_{\Gamma_2}=\sigma_z \psi_{\Gamma_2}$.
With a  suitable SU(2) gauge transformation on the spinons $F\rightarrow e^{i\mathbf{n\cdot}\boldsymbol{\tau}}F$,
one can choose a gauge in which $\Delta_{1\alpha}=0$, and the
hybridization $V_{1\alpha}$ is real. 
This hybridization in the first (singlet) channel then locks the U(1)
gauge of the spinons to the electrons, so that they have the same charge. 

We shall focus on the situation in which $J_{2}$ is smaller than
$J_{1}$ and seek superconducting solutions in which the normal state
$V_{1\alpha }=V$ is uniform and $\Delta_{2\alpha }$ are non-zero, leading to the gap equation
\begin{eqnarray}\label{}
\frac{    
\Delta_{2\alpha}}{J_{2}}
=-\frac{1}{N_s}\sum_{\mathbf{k}}\left\langle 
    f_{-\mathbf{k}\alpha}(i\sigma_{y})\sigma_{y}  \biggl (\sqrt{1-\zeta^{2}} + 
i {\zeta} \alpha p^{z}_{\bk }\biggr )
    c_{\mathbf{k}\alpha}\right\rangle \cr \notag
\end{eqnarray}
with the constraint $n_{f\alpha }=\frac{1}{N_{s}}\sum_{\mathbf{k}}\left\langle
f_{\mathbf{k}\alpha}^{\dagger}f_{\mathbf{k}\alpha}\right\rangle
=1$. We see that there are two types of
superconducting solution: 
\begin{enumerate}
\item a sublattice-even solution where $\Delta_{2\alpha
}=\Delta_2$ and the triplet pairing  term $\langle f_{-\bk \alpha} \sqrt{1-\zeta^2} c_{\bk \alpha }\rangle$
is an even function of $\bk $. 
This solution is then necessarily antisymmetric in the band indices -
an inter-band gap function without a Cooper instability, that will only develop for $J_{2}$ above a critical value. 

\item a staggered gap 
where $\Delta_{2\alpha }\propto  \alpha \Delta_{2}$ is odd on the
two sublattices and $\langle f_{-\bk \alpha} \zeta p_\bk^z c_{\bk
\alpha }\rangle$ is an odd function of momentum. This solution has a
Cooper instability. 
\end{enumerate}

Below the superconducting $T_c$ (Figure \eqref{fig:Temperature-dependence-of}), we find a sublattice-odd mean-field ground state
$V_{1A}=V_{1B},\Delta_{2A}=-\Delta_{2B},V_{2\alpha}=0$ corresponding
to an odd-parity triplet superconductor. 
\begin{figure}
\includegraphics[scale=0.75]{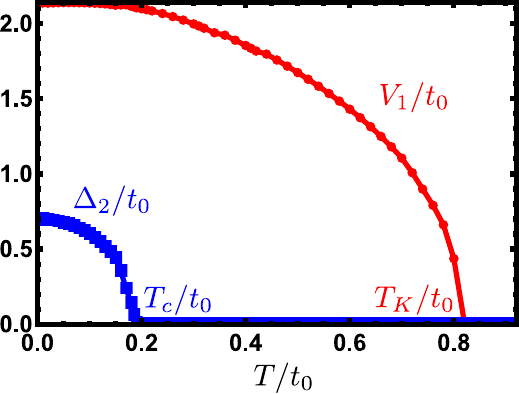}\caption{Temperature
dependence of order parameters. Calculations were made
with parameters  $J_{K1}=2.67t_0,J_{K2}=6.67t_0,\mu=2t_0,t_1=0.67t_0,\zeta=0.995$. \label{fig:Temperature-dependence-of}}
\end{figure}
The product of the order parameters in the two channels corresponds to a composite pairing operator\cite{coleman1999}
\begin{align}
\Psi=\langle V_{1\alpha} \Delta_{2\alpha} \rangle \propto \sum_{\mathbf{k},j,\alpha} \langle \psi_{2,-\mathbf{k}\alpha} \sigma_z (\boldsymbol{\sigma}i\sigma_y)\psi_{1\mathbf{k}\alpha} \cdot \boldsymbol{S}_{j\alpha} \rangle, \notag
\end{align}
which transforms as the $S_y=0$ component of a spin-triplet due to the non-trivial form-factor between the two channels $\sum_\alpha \alpha \langle \psi_{2,-\mathbf{k}\alpha}\sigma_z (i\sigma_y)\psi_{1\mathbf{k}\alpha}\rangle \sim p_\bk^z \sigma_y$.

In the Hund's-Kondo mechanism we present, the pairing symmetry is determined by the symmetry of the strongly-correlated \emph{atomic} states. In this demonstration, the pair potential transforms according to the $B_{3u}$ representation of $D_{2h}$, consistent with the irrep of the pre-formed pairs in the Hund's-coupled moment $|f^{2}\rangle\langle f^0|\sim \Phi_1^\dagger\sigma_z\Phi_2^\dagger \sim \sigma_z\sigma_x p_\bk^z$. This provides generic constraints on the pairing symmetries from the atomic matrix elements $|f^{n+2}\rangle\langle f^n|$ which can be experimentally accessed by resonant inelastic X-ray scattering~\cite{kotani2001,kotani2011} and by Raman scattering~\citep{devereaux2007}. There are two other generic features of our pairing mechanism that are experimentally verifiable. The sublattice-odd real-space form-factor of the pair wavefunction may be detected using scanning Josephson interferometry~\cite{coleman2020}. The continuous change in f-valence at the superconducting transition due to Kondo hybridization in a second channel is detectable using low-temperature core-level spectroscopy and X-ray scattering.

The pair-potential in our demonstration has line-nodes corresponding to the intersection of $p_{\bf k}^z =0$ and the Fermi surfaces, and a pseudo-spin d-vector on the Fermi surface aligned along the $y$-axis. In UTe$_2$, the experimental evidence for point-like gap nodes can be reconciled by noting that spin-orbit coupling induces secondary order parameters with d-vectors like $\mathbf{d}_\bk=p_\bk^y \hat{z}$ in the same irreducible representation, which gap out the line nodes except where $p_\bk^y=0$. This results in the d-vector of the pair potential having a two-dimensional texture in momentum space, and points to one of the many ways one could engineer non-trivial topology in this system. In our simple demonstration, the gaps on the two Fermi surfaces are identical and the superconductor is topologically trivial because any time-reversal invariant $\mathbb{Z}_2$ topological index is doubled. In UBe$_{13}$, a possible extension of this model would replace $p_{\bk}^z$ by an f-wave form factor that transforms like $k_x k_y k_z$. However, the observation that superconductivity emerges directly from a local moment regime suggests that the coupling strengths in the two channels are similar $J_1\approx J_2$, so that $T_c \approx T_{K}$. In the high temperature A phase of UPt$_3$, there is strong evidence~\cite{joynt2002,strand2010} for an f-wave superconducting gap and $S_z=0$ spin-triplet pairing, readily captured by a variation of our model in which the local pairs have $S_z=0$ spin-structure and the orbital content differs by the f-wave form factor $\Gamma_2\sim \Gamma_1 \times (x^2-y^2)z$.



There is good reason to suspect that the Kondo effect is intimately linked with the superconductivity in these materials. Tunneling experiments~\citep{jiao2020}
in UTe$_{2}$ find a Fano lineshape in the differential conductance,
that is characteristic of cotunneling into a local moment. The electrical resistivity of UTe$_{2}$\citep{aoki2019,ran2019} and CeRh$_{2}$As$_{2}$\citep{khim2021} both show clear broad
maxima in the temperature dependence, putatively
signaling the onset of Kondo hybridization. The superconducting $T_c$ is well below the Kondo temperature - the local moments are not spectators to the development of triplet pairing correlations.

Our theory contrasts with many recent theoretical proposals~\citep{ishizuka2019,nevidomskyy2020,shishidou2021,yu2022,yu2022a,yarzhemsky2020,shaffer2022,mineev2022,machida2020,machida2020a,hu2021,chen2021a,cavanagh2021,schertenleib2021,mockli2021,ptok2021,skurativska2021,hafner2022,nogaki2021,nogaki2022a}
for the pairing symmetry of UTe$_{2}$ and CeRh$_{2}$As$_{2}$ (with the noted exception of ~\citep{hafner2022}), in which Kondo physics does not play a role. Instead, the common theme is to consider a Kramer's
doublet at each site, forming narrow dispersive bands, whose Fermi
surfaces are unstable to pairing by intersite magnetic exchange interactions.
The symmetry of the superconducting order parameter is then determined
either by the anisotropy of ferromagnetic fluctuations~\citep{ishizuka2021}
or by the anisotropy of the band spin-orbit coupling in presence of
isotropic ferromagnetic exchange~\citep{shishidou2021}.
We have alternately emphasized that the symmetric-spin pairing correlations ascribed to inter-site interactions, ferromagnetic or antiferromagnetic~\citep{hu2021,chen2021a}, are already present at the atomic level, driven directly by the largest energy scale in the system - atomic Coulomb repulsion.   

The conceptual appeal of a Hund's-Kondo pairing mechanism lies in its ability to harness the coherence of Kondo hybridization to 
couple pre-formed Hund's triplets into a superconducting condensate. Key to this framework is the common structural motif of local moments with a sublattice degree of freedom shared by the diverse set of heavy fermion superconductors in Table \ref{tab:Heavy-fermion-triplet},  that allows localized triplet pairs to overlap with odd-parity Cooper pairs on the Fermi surface.

\emph{Acknowledgments: }This work was supported by 
the U.S. National Science Foundation grant DMR-1830707 (TH,PC). 
We gratefully acknowledge discussions
with Steven Kivelson, Daniel Agterberg, Vidya Madhavan, Silke Buhler-Paschen, Yashar Komijani and Victor Drouin-Touchette.

\bibliography{tripletKondo}
\appendix
\section{Derivation of singlet-triplet Kondo model from mixed-valent description}\label{sec:mixedValence}
The goal of this section is to derive the two-channel Kondo model in Eq.~(3) of the main text starting from a minimal description of the valence fluctuations of the f-electrons. Our starting point is a periodic Anderson model $H=H_c+H_f+H_{cf}$ that
captures the conduction electron mediated valence fluctuations of the three low-lying f-states in Figure~3 of the main text - the f-electron vacuum $|f^0\rangle$,
a single f$^{1}$ Kramers doublet $|f^1_{\Gamma_1,\sigma}\rangle$ and a Hund's coupled high-spin f$^{2}$ singlet $|f^2\rangle$.
The conduction electrons are described by a tight-binding Hamiltonian of electrons hopping on a two-sublattice body-centered cubic lattice as in Eq.~(3) of the main text
\begin{eqnarray}
H_{c} &=& - \sum_{\mathbf{k}}c\dg_{\bk}
\bigl[(t_{0}\gamma_\mathbf{k}+\mu) +
t_{1}\gamma_\mathbf{k}\alpha^{x}\bigr]c_{\mathbf{k}}
\end{eqnarray}
where $c\dg_{\bk }\equiv c\dg_{\bk \alpha \sigma }$ creates an electron of wavevector
$\bk $ on sublattice $\alpha =\pm 1$ with spin component $\sigma^{z}=\sigma$,
$t_0$ and $t_{1}$ are the intra- and inter-sublattice hopping integrals, respectively,
$({\alpha}^{x},{\alpha}^{y},{\alpha}^{z}) $ are the sublattice Pauli matrices and
$\gamma_\mathbf{k}=8\cos\frac{k_{x}}{2}\cos\frac{k_{y}}{2}\cos\frac{k_{z}}{2}$ 
is the nearest neighbor form-factor that is invariant under the D$_{2h}$ point group.
The energy of the f-electron configurations in the truncated Hilbert space is given by
\begin{eqnarray}
H_{f}&=& \sum_{\br} ( E_{2}X_{22} (\br)+ E_{0}X_{00} (\br) + E_{1}X_{\sigma \sigma } (\br))
\end{eqnarray}
where $X_{00}=|f^0\rangle\langle f^0|,X_{\sigma\sigma}=|f^1_{\Gamma_1,\sigma}\rangle\langle f^1_{\Gamma_1,\sigma}|, X_{22}=|f^2\rangle\langle f^2|$ are the Hubbard operators at each local moment site $\br\equiv (j,\alpha)$ defined by a unit cell $j$ at position $\bR_{j}$ and a sublattice $\alpha$.
The f$^2$ singlet wavefunction is $|f^{2}\rangle=f_{\Gamma_{2}s}^{\dagger}\left(\sigma_{z}i\sigma_{y}\right)_{ss^{\prime}}^{\dagger}f_{\Gamma_{1} s^{\prime}}^{\dagger}|0\rangle$ as in Figure~3 of the main text.
Thus the matrix element $\langle f^{2}|f_{\Gamma_{2}\sigma}^{\dagger}|f_{\Gamma_{1}\sigma^{\prime}}^{1}\rangle\sim\left(\sigma_{z}i\sigma_{y}\right)_{\sigma^{\prime}\sigma}$ and the hybridization between the conduction and f electrons at each local moment site is given by 
\begin{eqnarray}\label{app1}
H_{cf} & =&
\tilde{V_{1}}\sum_{\br}\left(X_{\sigma0}^{\Gamma_{1}}(\br)
\psi^{\Gamma_{1}}_{\sigma} (\br) +\mathrm{H.c.}\right) \cr
 & +&\tilde{V_{2}}\sum_{\br}\left(X_{2\sigma^{\prime}}^{\Gamma_{2}}(\br)
 \left(\sigma_{z}i\sigma_{y}\right)_{\sigma^{\prime}\sigma}
\psi^{\Gamma_{2}}_{\sigma} (\br) +\mathrm{H.c.}\right)\label{eq:latticeHcf}
\end{eqnarray}
where $X_{\sigma0}^{\Gamma_{1}}=|f_{\Gamma_{1}\sigma}^{1}\rangle\langle0|$
and $X_{2\sigma^{\prime}}^{\Gamma_{2}}=|f^{2}\rangle\langle f_{\Gamma_{1}\sigma^{\prime}}^{1}\vert$
are the projected f-electron creation operators
and $\psi_{\sigma}^{\Gamma}(\br)$ is a conduction electron
annihilation operator at site $\br$
whose spatial wavefunction transforms according to the irreducible representation $\Gamma\in (\Gamma_{1},\Gamma_{2})$ of the site symmetry group.
The important point for our pairing
mechanism is that the transitions between $f^{0}\leftrightharpoons f^{1}$
and $f^{1}\leftrightharpoons f^{2}$ are of different
symmetry and therefore hybridize with conduction electrons in different representations,
leading to a two-channel periodic Anderson model.

The Bloch-wave representation of these local conduction electron operators is 
\begin{equation}\label{}
\psi^{\Gamma}_{\bk\alpha\sigma }=\frac{1}{\sqrt{N_s}}\sum_{j} \psi_{ \sigma }^{\Gamma} (\br) e^{-i\bk\cdot {\bR}_{j}}
\end{equation}
where $N_s$ is the number of sites, ${\bR}_{j}$ is the location of the
unit cell, $\alpha =\pm 1$ is the sublattice index and $\sigma = \pm
1/2$ is the spin index. 
Due to spin-orbit coupling, the electron generically undergoes a spin-rotation as it escapes 
from a localized f-electron state to a conduction electron state with different orbital content. 
For simplicity, we shall adopt a model in which the conduction band Wannier
functions transform according to $\Gamma_{1}$ so that the valence fluctuations in the 
first channel to have the same symmetry as the conduction electron operators, 
$\psi^{\Gamma_1}_{\bk\alpha\sigma}=c_{\bk\alpha\sigma}$.
The effect of spin-orbit coupling in the model is then contained entirely in the second channel.
We take the case where the product of irreps $\Gamma_{2}\times\Gamma_{1}$ is isomorphic to $\sigma_{x}$
so that a symmetry-allowed form-factor in the second channel 
\begin{equation}\label{}
\psi^{\Gamma_{2}}_{\bk\alpha\sigma }=[\Phi^{\Gamma_{2}}_{\bk}]_{\alpha \sigma,\beta \sigma '} c_{\bk\beta\sigma '}
\end{equation}
is 
\begin{equation}\label{}
\Phi^{\Gamma_{2}}_{\bk}=i\sigma_x 
(\sqrt{1-\zeta^{2}}+i\zeta \alpha_{z} p_{\bk}^z)
, 
\end{equation}
where $\alpha_{i}$ and $\sigma_{i}$ are Pauli matrices in sublattice
and spin space, respectively, while $ p_\bk^z $ is the nearest neighbor crystal harmonic that
transforms like $z$. 
The $i\sigma_x$ form factor in $\Phi^{\Gamma_{2}}_{\bk}$ is a consequence of spin-orbit coupling - 
and the offset from the inversion center represented by $\zeta$ enables a linear coupling of even-parity spin and an odd-in-momentum form-factor $p_\bk^z$.
Note that the inversion interchanges the sublattices
so that $\alpha_{z} p_\bk^z $ is even parity. 
The construction is motivated by the structure of UTe$_{2}$ (see Figure 1 of the main text) but is
readily generalized. This leads to the Anderson model,
$H=H_{c}+H_{f}+H_{cf}$, where $H_{c}$ and $H_f$ describes the conduction
and atomic degrees of freedom respectively, and 
\begin{widetext}
\begin{align}
H_{cf} & =\tilde{V_{1}}\sum_{j\bk\alpha\sigma }\left(X_{\sigma 0}^{\Gamma_{1}} (j,\alpha ) c_{\bk\alpha\sigma }e^{i\bk\cdot {\bR}_{j\alpha}}+\mathrm{H.c.}\right)\nonumber \\
 & +\tilde{V_{2}}\sum_{jk\alpha\sigma\sigma '}\left(X_{2\sigma' }^{\Gamma_{2}}
 (j,\alpha ) 
 \biggl [\bigl (\sqrt{1-\zeta^{2}}+i\zeta \alpha_{z} p_{\bk}^z\bigr )
( {\bf d}\cdot \boldsymbol{\sigma })i\sigma_{y}\biggr ]_{\alpha \sigma' ,\beta \sigma}c_{\bk\beta\sigma}e^{i\bk\cdot \bR _{j }}+\mathrm{H.c.}\right).\label{eq:tripletcf}
\end{align}
describes the hybridization between the levels, where the d-vector
${\bf d}= \mathbf{\hat y}$ points in the y direction. 
The triplet hybridization appearing here has two terms - a spatially
symmetric term proportional to $\sqrt{1-\zeta^{2}}$ that is sublattice
uniform, and a second sublattice-odd term proportional to $\zeta \alpha_{z}$ which is spatially odd-parity. 
This odd-parity coupling to momentum is explicitly dependent on the displacement of the
center of symmetry away from the magnetic ions. This term 
is symmetry-equivalent to the antisymmetric spin-orbit
coupling, or ``staggered Rashba coupling'' discussed in the context of non-centrosymmetric superconductors~\citep{bauer2012}
and more recently, locally-noncentrosymmetric superconductors such as 
UTe$_{2}$ and CeRh$_{2}$As$_{2}$~\citep{sigrist2014,ishizuka2021}. 
The valence fluctuations can now be integrated out by a Schrieffer-Wolff
transformation~\cite{Schrieffer66,Coqblin69} which decouples the low-energy sector
$e^{iS}(H_{0}+H_{cf})e^{-iS}$. To leading order in the hybridization,
we achieve this decoupling by setting $i[S,H_{0}] = -H_{cf}$, which gives
\begin{eqnarray}\label{}
S= -i \sum_{\br}\bigg(\frac{\tilde{V_{1}}}{E_{1}-E_{0}} X_{\sigma 0}^{\Gamma_{1}} (\br) \psi_{\sigma }^{\Gamma_{1}} (\br)+\frac{\tilde{V_{2}}}{E_{2}-E_{1}} X_{2\sigma' }^{\Gamma_{2}} (\sigma_{z}i\sigma_{y})_{\sigma '\sigma }\psi^{\Gamma_{2}} _{\sigma } (\br)-\mathrm{H.c.}\bigg)
\end{eqnarray}
where the energy of the conduction electrons has been neglected
relative to the ionization energies $E_{0}-E_{1}$ and
$E_{2}-E_{1}$. The resulting Hamiltonian is 
then given by $H = H_{c}+
H_{K}$, where $H_{K} =\frac{i}{2}[S,H_{cf}]$, which 
leads to the two-channel Kondo model introduced in Eq (3) of the main text
\begin{eqnarray}
H_{c}&=& - \sum_{\mathbf{k}}c\dg_{\bk
}
\bigl[(t_{0}\gamma_\bk+\mu) +
t_{1}\gamma_\bk\alpha^{x}\bigr]c_{\bk},\cr
H_{K}&=&  J_{1}\sum_{j\alpha}
\psi\dg _{1\alpha} (j)\boldsymbol{\sigma}\psi_{1\alpha}(j)
\cdot\mathbf{S}_{j\alpha}+  J_{2}\sum_{j\alpha}
\psi\dg _{2\alpha} (j)\sigma_{z}\boldsymbol{\sigma}\sigma_{z}\psi_{2\alpha}(j)
\cdot\mathbf{S}_{j\alpha}.
\end{eqnarray}
with $J_{K1}\sim\left|\tilde{V}_{1}\right|^{2}/ 2(E_{0}-E_{1}),\:J_{K2}\sim\left|\tilde{V}_{2}\right|^{2}/ 2(E_{2}-E_{1}),$ and the spin $\mathbf{S}=|f^{1s}\rangle\boldsymbol{\sigma}_{ss^{\prime}}\langle f^{1s^{\prime}}|/2$.
Note that it is the valence fluctuations into the Hund's-coupled high-spin
$f^2$ state that results in a non-trivial triplet Kondo coupling of the conduction electron
and local moment spin.
\end{widetext}

\section{Triplet Kondo interaction as an interband pair scattering term}\label{sec:pairpot}

The goal of this section is to show explicitly that the Hund's-driven triplet Kondo coupling (Eq (6) in the main text)
\begin{align}\label{eq:pairint}
H_{K2}= -J_{2}^{*}\sum_{j }
 (\psi_{2 j}\dg 
\sigma_{z} (-i\sigma_{y})f\dg_{j })
 (f_{j } (i\sigma_{y})\sigma_{z} 
\psi_{2j }) 
\end{align}
in our model is an attractive pair scattering interaction for triplet pairs on the Fermi surface, as indicated in Eq (8) of the main text
\begin{equation}
H^{*} =- J_{2}^{*} |u_{\bk }v_{\bk }|_{\rm FS}^{2}\sum_{\bk, \bk '}\Lambda_{\bk }\dg \Lambda_{\bk' }.
\end{equation}
First we transform to the eigenbasis of the conduction Hamiltonian $H_c=- \sum_{\mathbf{k}}c\dg_{\bk} \bigl[(t_{0}\gamma_\bk+\mu) + t_{1}\gamma_\bk\alpha^{x}\bigr]c_{\bk}=\sum_\bk \tilde{c}\dg_{\bk \eta} \epsilon_\eta \tilde{c}_{\bk\eta}$ with $\tilde{c}_{\bk\eta}=U_{\bk\eta\alpha} c_{\bk\alpha}$, where $\eta$ labels the two conduction bands and $U=(1+i\alpha_y)/\sqrt{2}$ is the unitary transform that diagonalizes $H_c$. We apply the same transformation on the spinons $\tilde{f}_{\bk\eta}=U_{\bk\eta\alpha} f_{\bk\alpha}$ to get 
\begin{align}
H_c+H_{K1}=&\sum_\bk \tilde{c}\dg_{\bk \eta} \epsilon_\eta \tilde{c}_{\bk\eta} -\lambda \tilde{f}\dg_{\bk \eta} \tilde{f}_{\bk \eta} \notag\\
			&\qquad + V_1 f\dg_{\bk \alpha}  \Phi_{1\alpha\beta} c_{\bk \beta} + \mathrm{h.c.} \notag\\
		=&\sum_\bk \tilde{c}\dg_{\bk \eta} \epsilon_\eta \tilde{c}_{\bk\eta} -\lambda \tilde{f}\dg_{\bk \eta} \tilde{f}_{\bk \eta} \notag\\
			&\qquad + V_1 \tilde{f}\dg_{\bk \eta} \Phi_{1\eta\eta^\prime} \tilde{c}_{\bk \eta^\prime} + \mathrm{h.c.}	
\end{align}
where $\Phi_{1\eta\eta^\prime}=U\dg_{\eta\alpha} \Phi_{1\alpha\beta} U_{\beta\eta^\prime}$ is the form-factor of the first channel in the band basis.
Next we diagonalize the singlet Kondo hybridization to get the heavy fermion band eigenstates 
\begin{align}
    a_{\bk\eta} = u_{\bk\eta} \tilde{f}_{\bk\eta} +v_{\bk\eta} \Phi_{1\bk\eta\eta} \tilde{c}_{\bk\eta}
\end{align}
where the projection into the conduction bands is justified by the observation that $V_1\ll D$, the conduction electron bandwidth, so that each heavy fermion band is almost exclusively derived from \emph{one} conduction band. 

We now rewrite the triplet Kondo coupling in \eqref{eq:pairint} as an attractive pair scattering term after inserting a resolution of the identity $\alpha_z^2=1$ to allow the fermions pairs to acquire a sublattice-odd form factor
\begin{align}\label{eq:Lamdda}
	H_{K2} &= -\frac{J_2^*}{N_s} \sum_{\bk\bk^\prime} b\dg_\bk b_{\bk^\prime} \notag \\
    b_{\bk} &= f_{-\bk\alpha}(i\sigma_y)\sigma_z \alpha_z\Phi_{2\bk\alpha\beta} c_{\bk\beta}
\end{align}
where $N_s$ is the number of unit cells, $b\dg_\bk$ creates a charge-2e spinon-conduction electron pair with relative momentum $2\bk$. To project this pairing interaction onto the heavy fermion bands at the Fermi surface, we invert the unitary transforms above to get
\begin{align}
    P_\eta f_{\bk\alpha} P_\eta &= U^\dagger_{\bk\alpha\eta} u_{\bk\eta} a_{\bk\eta} \notag \\
    P_\eta c_{\bk\alpha} P_\eta &= U^\dagger_{\bk\alpha\eta} \Phi^\dagger_{1\bk\eta\eta} v_{\bk\eta} a_{\bk\eta}
\end{align}
where $P_\eta$ projects into a given band and we insert these into \eqref{eq:Lamdda} to get 
\begin{align}
   P_\eta b_{\bk} P_\eta &= a_{-\bk\eta} (i\sigma_y) \mathbf{d}_{\bk\eta}\cdot\boldsymbol{\sigma} a_{\bk\eta} 
\end{align}
with the d-vector of the intra-band triplet pair given by 
\begin{align}
    \mathbf{d}_{\bk\eta}\cdot\boldsymbol{\sigma} = \sigma_z \left( U^\dagger_{-\bk\alpha\eta} (\alpha_z\Phi_{2\bk})_{\alpha\beta} U^\dagger_{\bk\beta\eta}\right) \left( u_{-\bk\eta} \Phi_{1\bk\eta\eta}^\dagger v_{\bk\eta} \right) \notag
\end{align}
Being a triplet-pair, the d-vector must inherit the full antisymmetry of the pair potential $d_{\bk}=-d_{-\bk}^T$. Since there is no spin-orbit coupling in either the conducting hopping or the singlet hybridization in our model, $u_{-\bk}=u_\bk, U_{-\bk}=U_{\bk}$ and for the choice of $\Phi_{1\bk}=1,\Phi_{2\bk}=i\sigma_{x}(\sqrt{1-\zeta^{2}} +i\zeta \alpha^{z}p^{z}_{\bk})$
\begin{align}
    d_{\bk\eta}\cdot\boldsymbol{\sigma} &= \sigma_z \left( u_{\bk\eta}  v_{\bk\eta} \right) \notag \\
	&\times \left( \sum_\alpha (U^\dagger_{\bk\alpha\eta})^2 (\alpha \sqrt{1-\zeta^{2}} + i\zeta \sigma_x p_\bk^z ) \right)  \label{eq:dveclong}\\
    &= C_{\bk\eta}\zeta \sigma_y p_\bk^z 
\end{align}
where $C_{\bk\eta} = -i \sum_\alpha u_{\bk\eta}v_{\bk\eta}(U^\dagger_{\bk\alpha\eta})^2$. The first term in the parentheses in \eqref{eq:dveclong} vanishes because inversion symmetry requires the band eigenstates to have equal support on both sublattices $(U^\dagger_{\bk A\eta})^2=(U^\dagger_{\bk B\eta})^2$. Note that without the $\alpha_z$ form-factor in \eqref{eq:Lamdda}, it is the second term that would vanish and the remaining intra-band d-vector $d_{\bk\eta} \sim \sqrt{1-\zeta^{2}} \hat{y}$ then has no antisymmetric component. The only pairs that would be scattered by the triplet Kondo coupling would have to be inter-band pairs anti-symmetric under band-interchange and unavailable at low energies.

Thus the triplet Kondo coupling takes the form indicated in Eq (8) of the main text describing scattering on heavy fermion pairs on each Fermi surface, 
\begin{align}
H^*_{\eta}=-J_2^* |u_\bk v_\bk|^2_{\rm FS} \sum_{\bk\bk^\prime} \Lambda\dg_\bk \Lambda_{\bk^\prime}
\end{align}
with $\Lambda_\bk= \frac{P_\eta b_\bk P_\eta}{u_\bk v_\bk} = a_{-\bk}(i\sigma_y) \zeta \sigma_y p^z_\bk a_\bk$.

\end{document}